# Spectrum conversion and pattern preservation of Airy beams in fractional systems with a dynamical harmonic-oscillator potential

Xiaoqin Bai[1], Juan Bai[1], Boris A. Malomed[3,4], and Rongcao Yang[*,1,2]

[1]*School of Physics and Electronics Engineering, Shanxi University, Taiyuan 030006, China*
[2]*Collaborative Innovation Center of Extreme Optics, Shanxi University, Taiyuan 030006, China*
[3]*Department of Physical Electronics, School of Electrical Engineering, Faculty of Engineering, Tel Aviv University, Tel Aviv 69978, Israel*
[4]*Instituto de Alta Investigación, Universidad de Tarapacá, Casilla 7D, Arica, Chile*[†]
* Corresponding author: sxdxyrc@sxu.edu.cn
[†]Sabbatical address

**Abstract:** We investigate the dynamics of optical Airy beams in the one-dimensional fractional Schrödinger equation with a harmonic-oscillator (HO) potential subjected to modulation along the propagation distance. Deriving general solutions for propagating beams and particular solutions for Airy waves with/without chirp, we study analytically the spectrum conversion and pattern preservation for the chirp-free and chirped Airy beams in the fractional system including the HO potential with moiré-lattice, hyperbolic-secant, and delta-functional modulation formats. For the HO-moiré-lattice potential, it is found that the chirp-free Airy beam experiences multiple spectrum conversions between the Airy and Gaussian patterns in the momentum space, preserving the Airy pattern in the coordinate space. The chirp magnitude of the chirped Airy beam determines whether the spectrum conversion occurs in the momentum space, and the splitting and evolution direction of the beam in the coordinate space. For the HO-hyperbolic-secant potential, the chirp-free Airy beam undergoes spectrum conversion and tunneling, with the positions of the spectrum conversion and tunneling significantly depending on parameters of the hyperbolic-secant potential; however, the spectrum conversion and pattern preservation of the chirped Airy beam occurs only under a certain relation of the chirp and parameters of the potential. In the case of the HO-delta-functional potential, the chirp-free Airy beam experiences abrupt spectrum conversion and a two-step spectrum shift; however, for the chirped Airy beam, the spectrum conversion is affected by the relation between the chirp and height of the potential. Effects of the fractional Lévy index on the spectrum conversion and pattern preservation of the Airy beams under the action of the three modulation patterns considered here are also explored in detail.

## 1. Introduction

Recently, fractional systems have drawn much a great deal of interest as fractional derivatives naturally emerge in dynamical models of complex physical systems [1]. In quantum mechanics, the fractional Schrödinger equation (FSE) is derived by means of the Feynman's path integration, replacing the Brownian path trajectories by Lévy-flight ones [2-4]. Considering the mathematical analogy between the Schrödinger equation in quantum mechanics and the paraxial equation for the paraxial beam propagation in optics, Longhi had proposed a physical design of the FSE with the help of aspherical optical cavities in 2015 [5]. Recently, Liu *et al*. reported the first experimental realization of optical FSE in the temporal domain, with an effective fractional group-velocity dispersion [6]. Generally, the dynamics of beams and modes governed by FSEs is a hot research topic in optics. Zhang *et al*. have reported the symmetric split and zigzag trajectories of Gaussian beams in the free space [7] and harmonic-oscillator (HO) potential [8], respectively. Later, they have found that Bloch oscillation and Zener tunneling can be suppressed in fractional systems [9]. In addition, the Rabi oscillations [10], Anderson localization [11], and soliton stability [12-15] in fractional system have also drawn much interest. Special types of the beam dynamics, such as the soliton generation from super-Gaussian beams [16], quasi-stable propagation of necklace patterns [17], oscillations and autofocusing of Pearcey-Gaussian beams [18], and "superarrivals" of Gaussian wave packets [19] have also been reported in fractional systems. In particular, the dynamics of asymmetric Airy beams inspires special interests due to their unique properties, such as self-acceleration, self-healing and diffraction-free propagation [20-22]. Chen *et al.* studied periodic evolution and interaction of Airy beams in the FSE with an external potential [23]. Iomin addressed the acceleration of Airy beams in gravitational optics based on the FSE [24]. Bai *et al*. investigated modulated splitting and asymmetric conical diffraction of Airy beams in fractional system with a *PT*-symmetric potential [25]. So far, the study on the dynamics of Airy beams, especially for the chirped ones in fractional systems with the HO potential is still missing due to the difficulty of solving the corresponding FSE analytically.

The HO potentials play an important role in quantum mechanics and optics, and are commonly employed to describe traps in Bose-Einstein condensates [26] and optical manipulations [27, 28]. In optics, when the characteristic length of the response in a strongly nonlocal medium is much larger than the width of the optical beams, the nonlinear problem can be simplified to the linear one by transforming the strong nonlocality into the HO potential [29]. In usual systems with the HO potential, symmetric beams, such as Gaussians and solitons exhibit breathing dynamics, while the accelerating beams, such as Airy and Bessel ones, feature distinctive anharmonic propagation [27, 30]. However, in fractional system with the HO potential the evolution of Airy beams with/without chirps have not been investigated yet, especially from in an analytical form, which is a problem worthy exploring.

In this work, firstly, general analytical solutions describing the evolution of beams are derived for the fractional system with a time-dependent HO potential, which can be used to study the evolution of generic beams in a longitudinally modulated fractional system with the HO potential. Based on the general solutions, analytical solutions for the evolution of chirp-free and chirped Airy beams in the momentum and the coordinate spaces, respectively, are presented. Furthermore, the spectrum conversion and pattern preservation of Airy beams are investigated in detail in the momentum and the coordinate spaces for the HO potential combined with longitudinal moiré lattice, hyperbolic-secant, or delta-functional potential in the fractional system.

## 2. The theoretical model and analytical solutions

### *2.1 The theoretical model*

The paraxial evolution of optical beams in the fractional diffraction system with a modulated external potential is governed by the variable-coefficient FSE [8,31]

$$i\frac{\partial \Psi(x,z)}{\partial z} - \frac{1}{2}\left(-\frac{\partial^2}{\partial x^2}\right)^{\frac{\alpha}{2}} \Psi(x,z) + V(x,z)\Psi = 0, \qquad (1)$$

where $\Psi(x,z)$ is the beam envelope, $x$ and $z$ are the normalized transverse and longitudinal coordinates, respectively, $\alpha\,(0<\alpha\leq 2)$ is the Lévy index [6], and $V(x,z)$ denotes the modulated external potential. To explore effects of the fractional diffraction and longitudinally modulated HO potential analytically, we focus on the limit case of $\alpha=1$ and $V(x,z)=-\frac{1}{2}v(z)x^2$ in the subsequent consideration. The case of $\alpha\neq 1$ should be considered by means of numerical methods.

For $\alpha=1$, the Fourier transform (FT) casts Eq. (1) in the following form in the momentum space:

$$i\frac{\partial \hat{\Psi}(k,z)}{\partial z} + \frac{v(z)}{2}\frac{\partial^2 \hat{\Psi}(k,z)}{\partial k^2} + fk\hat{\Psi}(k,z) = 0, \qquad (2)$$

where $\hat{\Psi}(k,z)$ is the FT of $\Psi(x,z)$, $k$ is the spatial frequency (wavenumber), and $f=-1/2\,\mathrm{sgn}(k)$. Obviously, Eq. (2) itself looks as a linear Schrödinger equation with a variable diffraction coefficient $v(z)$.

### *2.2 The analytical method and general solution*

Drawing on the solving method reported in [32], we introduce new variables

$$\xi(k,z) = k - f\int_0^z \varepsilon v(\varepsilon)d\varepsilon,\ \tau(z) = \int_0^z v(\varepsilon)d\varepsilon, \tag{3}$$

and adopt $\hat{\Psi}(k,z)$ in the form of

$$\hat{\Psi}(k,z) = \phi(\xi,\tau)F(k,z), \tag{4}$$

where $F(k,z)$ is also assumed to be a solution of Eq. (2). Substituting ansatz (4) in Eq. (2) and defining $\phi(\xi,\tau)$ as a solution of the free-space Schrödinger equation

$$i\frac{\partial \phi(\xi,\tau)}{\partial \tau} + \frac{1}{2}\frac{\partial^2 \phi(\xi,\tau)}{\partial \xi^2} = 0. \tag{5}$$

We arrive at an equation for $F(k,z)$:

$$i\xi_z' F(k,z) + v(z)\frac{\partial F(k,z)}{\partial k} = 0, \tag{6}$$

where $\xi_z' = -fzv(z)$ represents the derivative of $\xi(k,z)$ with respect to $z$.

From Eqs. (5) and (6), it is easy to obtain solutions

$$\phi(\xi,\tau) = \frac{1}{\sqrt{2\pi i\tau}}\int_{-\infty}^{\infty}\phi(s,0)\exp\left[\frac{i}{2\tau}(\xi-s)^2\right]ds, \tag{7a}$$

$$F(k,z) = \exp\left[ifkz - \frac{if^2}{2}\int_0^z \varepsilon^2 v(\varepsilon)d\varepsilon\right]. \tag{7b}$$

Then, it follows from Eqs. (3) and (7b) that $\tau(0) = 0$ and $F(k,0) = 1$. Thus, according to Eq. (4), we obtain $\phi(\xi,0) = \hat{\Psi}(k,0)$. Then, one can finally obtain an explicit analytical solution of the underlying equation (2):

$$\hat{\Psi}(k,z) = \frac{1}{\sqrt{2\pi i\tau(z)}}\exp\left[ifkz - \frac{if^2}{2}\int_0^z \varepsilon^2 v(\varepsilon)d\varepsilon\right]\int_{-\infty}^{\infty}\hat{\Psi}(s,0)\exp\left\{\frac{i}{2\tau(z)}\left[k - f\int_0^z \varepsilon v(\varepsilon)d\varepsilon - s\right]^2\right\}ds. \tag{8}$$

Note that the analytical solution (8) is meaningful for $\tau(z) \neq 0$. In the case of $\tau(z) = 0$, assuming $z' = z - z_j$, $k' = k - f\int_0^{z_j} \varepsilon v(\varepsilon)d\varepsilon$, where $\tau(z_j) = 0$ at singular position $z_j$ ($j = 0,1,2,\ldots,$ and $z_0 = 0$), it is easy to find $\phi(\xi,\tau(z)|z = z_j) = \phi\left[k - f\int_0^{z_j} \varepsilon v(\varepsilon)d\varepsilon, \tau(z')|z' = 0\right]$.

Thus, employing Eq. (3) and the relation $\phi(\xi,0)=\hat{\Psi}(k,0)$ derived above, one obtains

$$\phi(\xi,\tau)=\phi\left[\xi(k,z),0\right]=\phi(k',z'=0)=\hat{\Psi}(k',0). \tag{9}$$

It follows from Eq. (4) that the field distribution in the momentum space can be written, when $\tau(z)=0$, as

$$\hat{\Psi}(k,z)=\hat{\Psi}\left[k-f\int_0^z \varepsilon v(\varepsilon)d\varepsilon, 0\right]\exp\left[ifkz-\frac{if^2}{2}\int_0^z \varepsilon^2 v(\varepsilon)d\varepsilon\right]. \tag{10}$$

Combining Eqs. (8) and (10), one can obtain

$$\hat{\Psi}(k,z)=\begin{cases}\hat{\Psi}\left[k-f\varsigma(z),0\right]\exp\left[ifkz-\dfrac{if^2\eta(z)}{2}\right], & z=z_j\ (j=0,1,2,...) \\ \dfrac{1}{\sqrt{2\pi i\tau(z)}}\exp\left[ifkz-\dfrac{if^2\eta(z)}{2}\right]\int_{-\infty}^{+\infty}\hat{\Psi}(s,0)\exp\left\{\dfrac{i}{2\tau(z)}\left[k-f\varsigma(z)-s\right]^2\right\}ds, & z\neq z_j\end{cases}, \tag{11}$$

where $\varsigma(z)=\int_0^z \varepsilon v(\varepsilon)d\varepsilon$, $\eta(z)=\int_0^z \varepsilon^2 v(\varepsilon)d\varepsilon$.

Taking the inverse FT of Eq. (11), one can obtain the analytical solution in the coordinate space:

$$\Psi(x,z)=\begin{cases}\Psi(x+fz,0)\exp\left[if(x+fz)\varsigma(z)-\dfrac{if^2\eta(z)}{2}\right], & z=z_j\ (j=0,1,2,...) \\ \Psi(x+fz,0)\exp\left[if(x+fz)\varsigma(z)-\dfrac{if^2\eta(z)}{2}\right]\exp\left[-\dfrac{i\tau(z)}{2}(x+fz)^2\right], & z\neq z_j\end{cases}. \tag{12}$$

In fact, due to $\tau(z_j)=0$, the analytical solution (12) can be cast in the following one:

$$\Psi(x,z)=\Psi(x+fz,0)\exp\left[if(x+fz)\varsigma(z)-\frac{if^2\eta(z)}{2}\right]\exp\left[-\frac{i\tau(z)}{2}(x+fz)^2\right]. \tag{13}$$

For the coordinate function $v(z)$, according to the above expressions of $\varsigma(z), \eta(z)$ and $\tau(z)$, it is found from Eq. (13) that $\varsigma(z), \eta(z)$ and $\tau(z)$ affect only the phase of field $\Psi(x,z)$. This means the amplitude evolution of the beam maintains the initial pattern for a given longitudinal modulation. For the special case of constant coefficient $v(z)$, solution (11) in the momentum space can be reduced into the that studied in Ref. [8], while the general analytical solution (13) in the coordinate space has not been reported in Ref. [8]. It is relevant to note that Eqs. (11) and (13) possess general analytical solutions in the momentum and the coordinate spaces, without any restriction imposed on the form of the longitudinal modulation function $v(z)$. In this paper, we attempt to seek for an analytical Airy-wave solution of Eqs. (1) and (2) for $\alpha = 1$ and explore the respective dynamics of Airy beams in the fractional diffraction system with a longitudinally modulated HO potential.

### 2.3 An analytical chirp-free Airy solution

Firstly, we consider an initial obliquely launched chirp-free Airy wave,

$$\Psi(x,0) = \mathrm{Ai}(x)\exp(ax)\exp(iCx), \tag{14}$$

where $\mathrm{Ai}(\cdot)$ is the Airy function, $a > 0$ is the exponential truncation factor, and $C$ determined the initial launch angle. To present the generic case, we set $a = 0.1$ throughout the paper.

Substituting the FT of expression (14) in Eqs. (11) and (13), we derive analytical solutions of Eqs. (1) and (2) for the chirp-free Airy beam in the momentum and the coordinate spaces, respectively:

$$\hat{\Psi}(k,z) = \begin{cases} \exp\left\{-a\left[k - f\varsigma(z) - C\right]^2\right\}\exp\left\{\frac{i}{3}\left(\left[k - f\varsigma(z) - C\right]^3 - 3a^2\left[k - f\varsigma(z) - C\right] - ia^3\right)\right\} \\ \times \exp\left[ifkz - \frac{if^2\eta(z)}{2}\right], \quad z = z_j \\[6pt] \sqrt{\frac{2\pi}{i\tau(z)}}\mathrm{Ai}\left[-\frac{k - f\varsigma(z) - C + ia}{\tau(z)} - \frac{1}{4\tau(z)^2}\right]\exp\left[ifkz - \frac{if^2\eta(z)}{2}\right] \\ \times \exp\left\{\frac{i\left[k - f\varsigma(z) - C + ia\right]^2}{2\tau(z)} + \frac{i\left[k - f\varsigma(z) - C + ia\right]}{2\tau(z)^2} + \frac{i}{12\tau(z)^3}\right\}, \quad z \neq z_j \end{cases} \tag{15a}$$

$$\Psi(x,z) = \mathrm{Ai}(x + fz)\exp\left[-\frac{i\tau(z)}{2}(x + fz)^2\right]\exp\left\{\left[if\varsigma(z) + iC + a\right](x + fz) - \frac{if^2\eta(z)}{2}\right\}. \tag{15b}$$

Note that, for the chirp-free Airy beam, Eq. (15) demonstrates spectrum conversion between the Airy and Gaussian patterns in the momentum space in the course of the propagation, while the Airy pattern is preserved in the coordinate space.

### 2.4 The analytical chirped Airy solution

Considering the important role of chirp in optics and noteworthy features of the Airy-beam dynamics caused by the chirp, such as asymmetric splitting [25] and inversion transformation [33], it is necessary to study the effect of the chirp on the evolution of the Airy beams. For the initial obliquely launched chirped Airy wave,

$$\Psi_c(x,0) = \mathrm{Ai}(x)\exp(ax)\exp(iCx)\exp(ibx^2), \tag{16}$$

where $b$ is the chirp parameter, following the steps demonstrated above, one can obtain evolution solutions of Eqs. (1) and (2) by inserting the FT of the input (16),

$$\hat{\Psi}_c(k,0) = \sqrt{i\frac{\pi}{b}}\,\mathrm{Ai}\!\left(\frac{k-C+ia}{2b} - \frac{1}{16b^2}\right)\exp\!\left[-\frac{i(k-C)^2}{4b} + \frac{i(k-C)}{8b^2} + \frac{a(k-C)}{2b} - \frac{i}{96b^3} + \frac{ia^2}{4b} - \frac{a}{8b^2}\right], \tag{17}$$

into Eqs. (11) and (13). Thus, the solution for the chirped Airy beam in the momentum and the coordinate spaces can be derived as

$$\hat{\Psi}_c(k,z) = \begin{cases} \sqrt{i\dfrac{\pi}{b}}\,\mathrm{Ai}\!\left[\dfrac{k-f\varsigma(z)-C+ia}{2b} - \dfrac{1}{16b^2}\right]\exp\!\left[ifkz - \dfrac{if^2\eta(z)}{2}\right] \\ \quad \times \exp\!\left\{-\dfrac{i[k-f\varsigma(z)-C+ia]^2}{4b} + \dfrac{i[k-f\varsigma(z)-C+ia]}{8b^2} - \dfrac{i}{96b^3}\right\},\quad z=z_j \\[1em] \sqrt{\dfrac{2\pi}{i[\tau(z)-2b]}}\,\mathrm{Ai}\!\left\{-\dfrac{k-f\varsigma(z)-C+ia}{\tau(z)-2b} - \dfrac{1}{4[\tau(z)-2b]^2}\right\}\exp\!\left[ifkz - \dfrac{if^2\eta(z)}{2}\right] \\ \quad \times \exp\!\left\{\dfrac{i[k-f\varsigma(z)-C+ia]^2}{2[\tau(z)-2b]} + \dfrac{i[k-f\varsigma(z)-C+ia]}{2[\tau(z)-2b]^2} + \dfrac{i}{12[\tau(z)-2b]^3}\right\},\quad z\neq z_j \end{cases} \tag{18a}$$

$$\Psi_c(x,z) = \mathrm{Ai}(x+fz)\exp\!\left\{\left[-\frac{i\tau(z)}{2}+ib\right](x+fz)^2\right\}\exp\!\left\{[if\varsigma(z)+iC+a](x+fz)-\frac{if^2\eta(z)}{2}\right\}. \tag{18b}$$

It should be noted that solution (18a) for the chirped Airy beam in the momentum space cannot be straightforwardly reduced to the case without the chirp by setting $b=0$, as $b$ appears in the denominator in Eq. (17). Moreover, the two expressions for the solution given by Eq. (18a) with $z=z_j$ and $z\neq z_j$ cannot be merged into one because the condition of $b\neq 0$ is necessary in the former expression, but it is not required in the latter one. Furthermore, it can be seen from Eq. (18a) that, in the momentum space, the spectrum conversion between the Airy and Gaussian patterns does not occur in the course of the propagation, only Airy patterns being formed, which is different from the case of the chirp-free Airy beam. However, in the coordinate space, solution (18b) for the chirped Airy beam can be reduced to Eq. (15b) for the case without the chirp. In addition, it is noted that condition $\tau(z)\neq 2b$ is required for the validity of Eq. (18a). In the special case of $\tau(z)=2b$, the analytical solutions at singular positions for the chirped Airy beam in the momentum and the coordinate spaces take the following form:

$$\hat{\Psi}_c(k,z) = \exp\left\{-a[k - f\varsigma(z) - C]^2\right\}\exp\left\{\frac{i}{3}\left([k - f\varsigma(z) - C]^3 - 3a^2[k - f\varsigma(z) - C] - ia^3\right)\right\}\exp\left[ifkz - \frac{if^2\eta(z)}{2}\right],$$

(19a)

$$\Psi_c(x,z) = \mathrm{Ai}(x + fz)\exp\left\{[if\varsigma(z) + iC + a](x + fz) - \frac{if^2\eta(z)}{2}\right\}.$$ (19b)

This means that in the special case of $\tau(z) = 2b$, the chirped Airy beam is converted into the Gaussian beam at the singular positions in the momentum space, and the Airy pattern is preserved in the coordinate space.

## 3. Results and discussion

The above analytical solutions demonstrate that the external potential $V(x,z)$ has great influence on the evolution of the beams in both the momentum and coordinate spaces, as $\varsigma(z)$, $\eta(z)$ and $\tau(z)$ depend on the form of $v(z)$. Below we explore the evolution of the chirp-free and chirped Airy beams in the fractional diffraction system with three types of HO potentials.

### 3.1 The HO-moiré lattice potential

In view of the important effect of moiré lattices on the wave localization and diffraction suppressing, and their potential applications to artistic design, architecture, image processing and so on [34,35], we first consider a longitudinal moiré lattice modulating the HO potential, namely, $v(z) = d_0^2\left[A_1\cos\left(\frac{2\pi}{\omega_1}z\right) + A_2\cos\left(\frac{2\pi}{\omega_2}z\right)\right]$ [36, 37], in which $\omega_1 = \omega(1+\varpi)$, $\omega_2 = \omega(1-\varpi)$, $\varpi \ll 1$, as shown in Fig. 1. Clearly, the potential is parabolic (HO) along the transverse coordinate, while following the form of the moiré lattice along the propagation distance, whose modulation depth and period strongly depend on the values of $\omega_1$, $\omega_2$ and $A_2/A_1$, as plotted in Fig. 1(b).

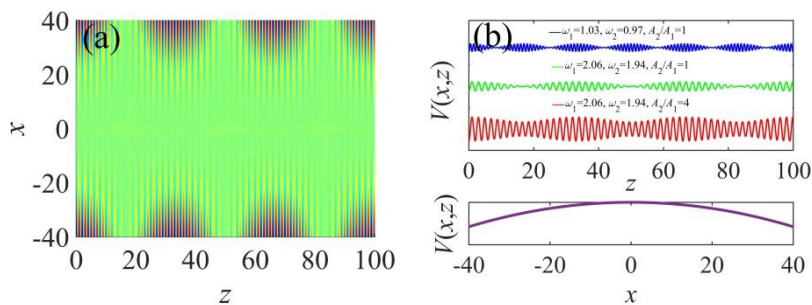

Fig. 1 The HO-moiré lattice potential $V(x,z)$: (a) the spatial distribution; (b) (top) side views along the $z$ axis with different $\omega_1$, $\omega_2$ and $A_2/A_1$, and (bottom) side view along the $x$ axis.

In accordance with Eq. (15a), for the dynamical HO-moiré lattice potential, the chirp-free Airy beam exhibits two different evolution patterns with spectrum conversions in the momentum space in the course of the propagation. Namely, the beam features the Gaussian shape at $z = z_j$ [$z_j$ satisfies the condition $\sin(2\pi z_j/\omega_1)/\sin(2\pi z_j/\omega_2) = -A_2\omega_2/A_1\omega_1$], while converting into an Airy pattern at other propagation distances, as shown in Figs. 2 (a1) and (a2). Moreover, the beam undergoes the inversion with opposite acceleration after passing points $z = z_j$ in the momentum space, as the sign of $\tau(z)$ is opposite before and after $z = z_j$. In the coordinate space, the propagating beam preserves its Airy pattern, as seen from Figs. 2 (b1) and (b2). The analytical and numerically produced distributions are in good agreement with each other, as shown in Figs. 2 (a2) and (b2). The long-distance evolution of the beam is depicted in Figs. 2 (a3) and (b3). It is clearly seen that the beam exhibits oscillatory propagation in both the momentum and coordinate spaces due to the longitudinal modulation $v(z)$. Comparing Figs. 2 (a3) and 2 (b3), it is found that the beam propagates away from $x = 0$ in the coordinate space, when it travels close to $k = 0$ in the momentum space. This can be explained by Eq. (15), where velocities in the momentum and the coordinate spaces are given by $fzv(z)$ and $-f$, respectively. Seemingly, the beam experiences the action of a restoring force [8], whose sign is reversed when the beam crosses the axis of $k = 0$. Moreover, the interference between the beam lobes appears during the propagation in the coordinate space [see the inset of Fig. 2 (b3)], which is caused by the fact that the beam is driven by two opposite restoring forces when side lobes travel across the axis of $k = 0$ in the momentum space.

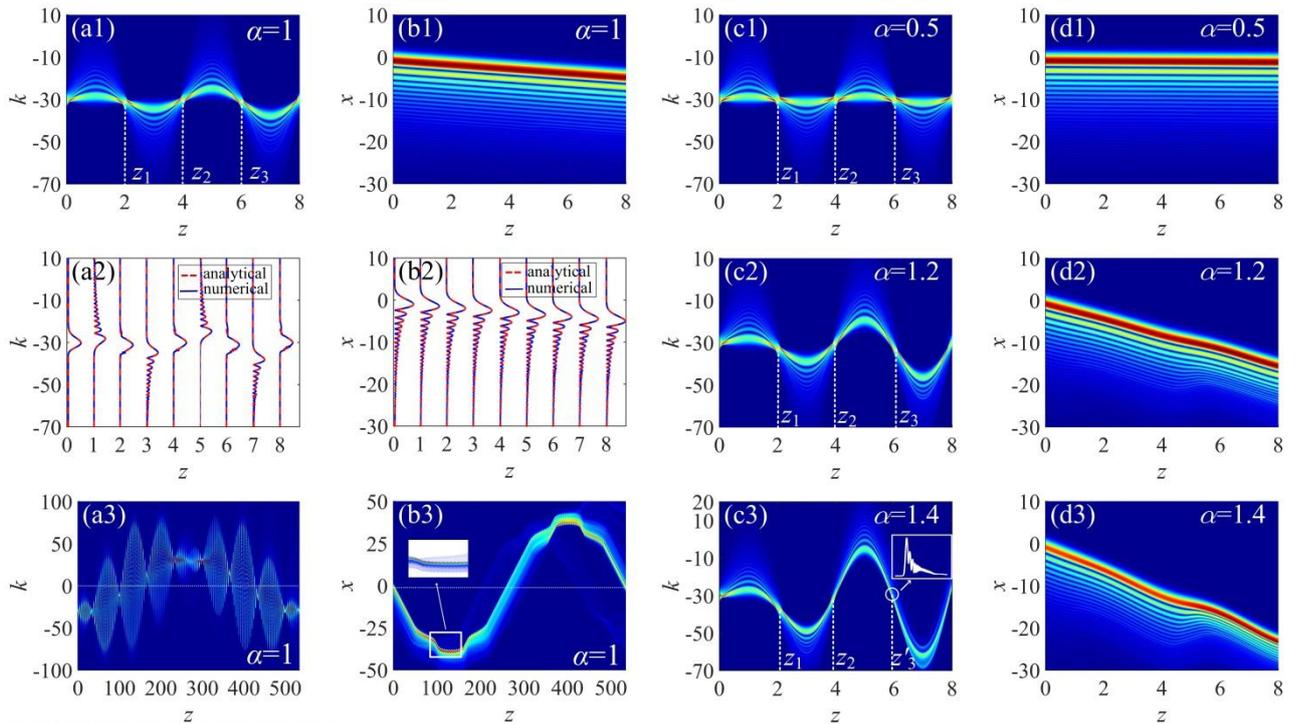

Fig. 2 The evolution of the chirp-free Airy beam in the momentum (the first and third columns) and the coordinate (the second and fourth columns) spaces. (a,b) For $\alpha = 1$: (a1,b1) for a short propagation distance; (a2,b2) amplitude profiles at different propagation distances, $z = 0,1,2,3,4,5,6,7,8$; (a3,b3) for a long propagation distance. (c,d) For $\alpha \neq 1$: (c1,d1) $\alpha = 0.5$; (c2, d2) $\alpha = 1.2$; (c3, d3) $\alpha = 1.4$. Other parameters are: $C = -30$, $d_0 = 1.2$, $A_1 = A_2 = 1$, $\omega_1 = 4.12$, $\omega_2 = 3.88$.

It is relevant to comment on the fact that such hallmarks of Airy beams as self-bending and self-acceleration are not observed in Fig. 2. This is because the actual trajectories of the Airy lobes, being determined by the external potential, are different from the self-bending ones in the free space [20,21]. Specifically, the trajectories in the coordinate and momentum spaces are determined, respectively, by expressions $x = -fz = \text{sgn}(k)z/2$ and $k = f\varsigma(z) + C$ with $\varsigma(z) = \int_0^z \varepsilon v(\varepsilon) d\varepsilon$, according to Eq. (15), which implies a linear trajectory in the coordinate space and an oscillatory one in the momentum space, under the action of the HO-moiré lattice potential. Specially, when $v(z) = 0$, the dynamics of the Airy beam governed by the FSE are reduced to that reported in Refs. [23,25], where the beams split symmetrically and propagate along straight lines in the coordinate space and follow the Gaussian distribution in the momentum space.

To explore the effect of Lévy index $\alpha$ on the spectrum conversion and pattern preservation, we numerically simulated the evolution of the chirp-free Airy beam for $\alpha = 0.5$, 1.2 and 1.4, as shown in Figs. 2 (c) and (d). It is observed that the spectrum conversion in the momentum space and pattern preservation in the coordinate space still take place for $\alpha < 1$. Further, $\alpha$ affects the oscillation amplitude of the Airy pattern in the momentum space [see Fig. 2 (c)] and the evolution direction of the beam in the coordinate space [see Fig. 2 (d)]. Moreover, the spectrum conversion gradually disappears with the increase of the propagation distance for larger $\alpha$, as shown in the inset of Fig. 2 (c3). Note that, in Fig. 2 (c3) the propagation distance, corresponding to $z_3$ in Figs. 2 (c1) and (c2), is labeled as $z'_3$, to distinguish it from the singular positions and indicate that the evolution of the beam in this range of the propagation distance follows the Airy pattern.

Furthermore, one can see from Eq. (15) that the evolution of the Airy beam strongly depends on parameters of the longitudinally modulated lattice. We take the modulation amplitudes $A_1$, $A_2$ and frequency $\omega$ (or $\omega_1$ and $\omega_2$) as examples to show the influence of the parameters of the longitudinal moiré lattice on the evolution of the chirp-free Airy beam in the case of $\alpha = 1$. As shown in Figs. 3 (a) and (b), the singular positions $z_j$ and oscillation amplitudes are significantly impacted in the momentum space, while only side lobes of the Airy pattern are slightly affected in the coordinate space.

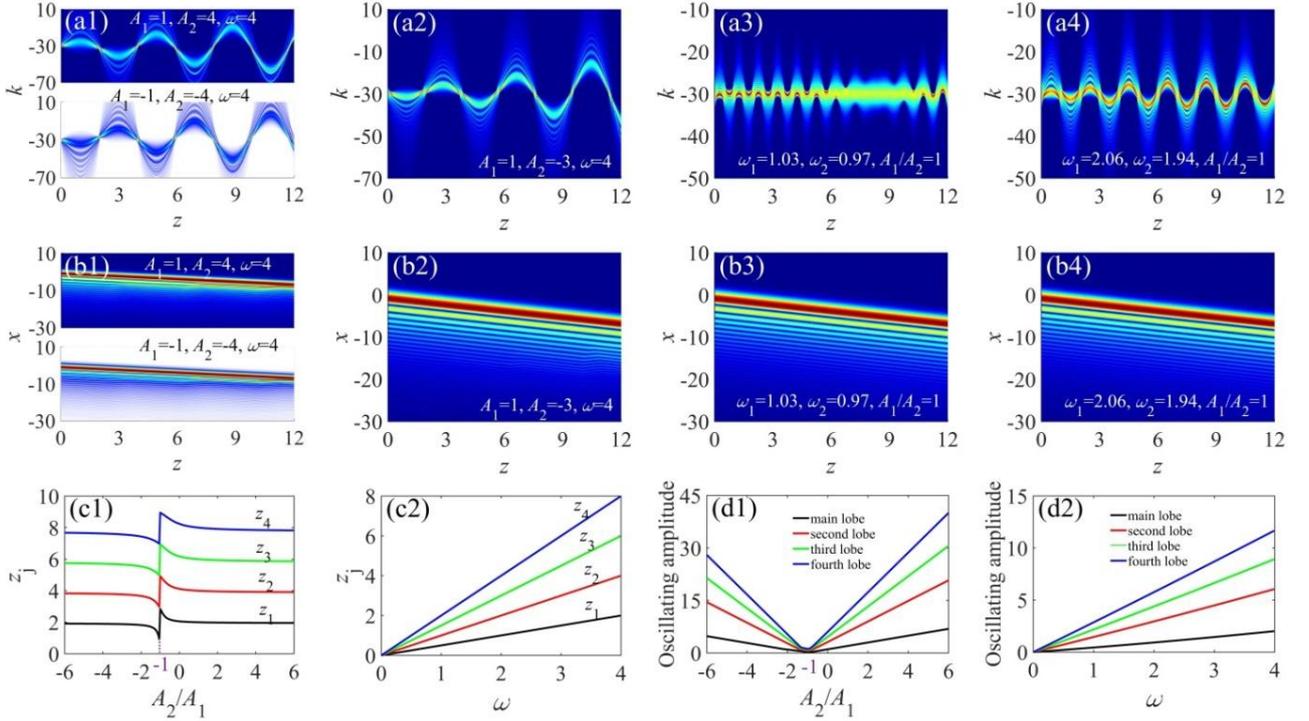

Fig. 3 The effect of the lattice parameters on the evolution of the chirp-free Airy beam in the momentum (first row) and the coordinate (second row) spaces for $\alpha = 1$: (a1,b1) $A_1 = 1$, $A_2 = 4$ and $A_1 = -1$, $A_2 = -4$ for $\omega = 4$; (a2,b2) $A_1 = 1$, $A_2 = -3$ for $\omega = 4$; (a3,b3) $\omega_1 = 1.03$, $\omega_2 = 0.97$ for $A_2/A_1 = 1$; (a4,b4) $\omega_1 = 2.06$, $\omega_2 = 1.94$

for $A_2/A_1 = 1$; (c1,c2) the first four non-zero $z_j$ and (d1,d2) the first oscillation amplitudes of the first four Airy lobes for different values of $A_2/A_1$ and $\omega$. Other parameters are the same as in Fig. 2.

In particular, when both $A_1$ and $A_2$ take opposite values, the evolution of the beam is mirrored about the axis of $k = C$ in the momentum space [see Fig. 3(a1)]. According to the constraint relationship demonstrated above, $\sin(2\pi z_j/\omega_1)/\sin(2\pi z_j/\omega_2) = -A_2\omega_2/A_1\omega_1$, the variation of singular positions $z_j$ with the change of $A_2/A_1$ and $\omega$ can be readily obtained [see Figs. 3(c1) and (c2)]. Figures 3(d1) and (d2) present the oscillation amplitudes of the Airy lobes varying with $A_2/A_1$ and $\omega$, respectively. Specially, the critical values of $z_j$ and oscillation amplitudes occur at $A_2/A_1 \approx -1$, and the positive ratio of $A_2/A_1$ leads to larger $z_j$ and oscillation amplitudes than its negative value, as shown in Figs. 3(c1) and (d1). Moreover, the oscillation amplitudes monotonously increase with $\omega$, as displayed in Figs. 3(c2) and (d2).

From the solutions given by Eqs. (18) and (19) for the chirped Airy beam, it is seen that the evolution of the chirped Airy beam strongly depends on the relation between $\tau(z) = \int_0^z v(\varepsilon)d\varepsilon$ and chirp parameter $b$. The evolution of the chirped Airy beam for the cases of $\tau(z) \neq 2b$ and $\tau(z) = 2b$ is shown in Figs. 4 and 5, respectively. It is found that, unlike the case without the chirp, when $\tau(z) \neq 2b$, the spectrum conversion between the Airy and Gaussian patterns does not occur and the beam keeps the form of the Airy pattern in the momentum space [see Figs. 4(a1) and (a2)], which can also be deduced from the expression in Eq. (18a).

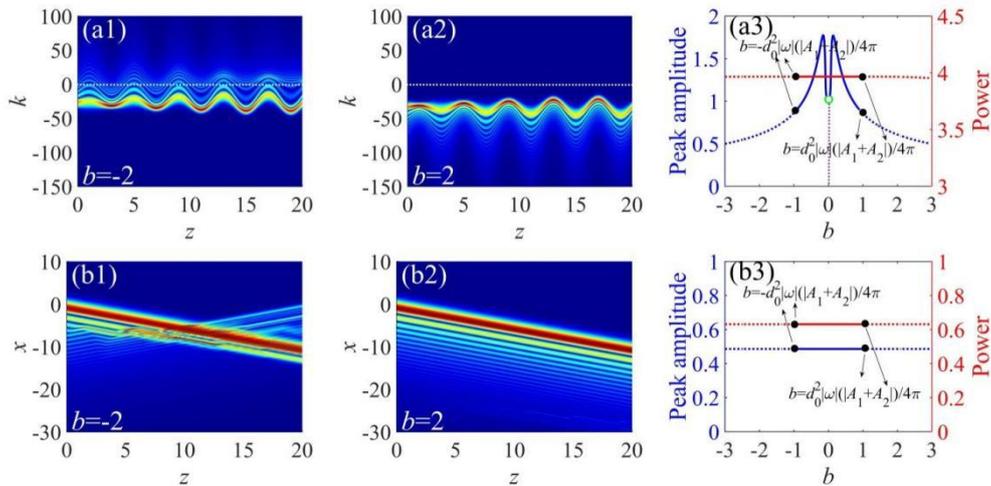

Fig. 4 The evolution of the chirped Airy beam in the momentum (top row) and the coordinate (bottom row) spaces for the case of $\tau(z) \neq 2b$ and $\alpha = 1$: (a1,b1) $b = -2$; (a2,b2) $b = 2$; (a3,b3) the variation of the peak amplitude and power at $z = 20$ with the change of chirp $b$ in the momentum and the coordinate spaces, respectively. Other parameters are the same as in Fig. 2.

Note that, in the case of negative chirp, there are some side lobes of the Airy pattern traveling across the axis of $k = 0$ in the momentum space, thus the beam is driven by two restoring forces in opposite directions, resulting in the fact that the Airy beam with negative chirp evolves into two parts with opposite velocities in the coordinate space [see Fig. 4(b1)]. However, for the case of positive chirp, as all lobes of the Airy pattern stay on the same side of axis $k = 0$, the beam always propagates keeping the Airy pattern in the coordinate space [see Fig. 4(b2)]. Figures 4(a3) and (b3) present the variation of the peak amplitude and power with chirp parameter $b$ at $z = 20$ in the momentum and the coordinate spaces, respectively. It is clearly seen that the peak amplitude increases at first, and then decreases with the increase of $|b|$, while the power remains unchanged in the momentum space, as shown in Fig. 4 (a3). On the other hand, both the peak amplitude and power stay constant with the variation of chirp $b$ in the coordinate space, as shown in Fig. 4(b3). In addition, we note that there are discontinuities at $b = 0$ and -0.007 (marked by the green circle; only one circle is seen as the distance between the two discontinuities is extremely small) for the peak amplitude in the momentum space. One discontinuity corresponds to the condition of $b \neq 0$, and the discontinuity at $b = -0.007$ corresponds to the condition $\tau(z=20)=2b$ in Eq. (18a).

For the case of $\tau(z) = 2b$, it is easy to deduce from Eqs. (18a) and (19a) that the spectrum conversion between the Airy and Gaussian patterns can happen for the chirped Airy beam in the momentum space. In fact, the condition $\tau(z) = 2b$ implies $\min[\tau(z)] \leq 2b \leq \max[\tau(z)]$. For the above longitudinal moiré lattice, it is easy to obtain $|b| \leq \dfrac{d_0^2 |\omega|(|A_1 + A_2|)}{4\pi}$ [see Fig. 5(a)]. Apparently, only when $|b| \leq \dfrac{d_0^2 |\omega|(|A_1 + A_2|)}{4\pi}$, the spectrum conversion can occur in the momentum space [see Fig. 5(b)], and the singular positions are $\tau(z_j) = 2b$, which are marked by pink circles in Fig.5(a). Nevertheless, the beam keeps the Airy shape with a certain velocity in the coordinate space

[see Fig. 5(c)], as demonstrated by Eqs. (18b) and (19b). Furthermore, for the chirped Airy beam, the increase of Lévy index $\alpha$ results in enhanced oscillations of the Airy pattern in the momentum space [see Figs. 5(d)] and a large change of the evolution velocity in the coordinate space [see Figs. 5(e)]. In particular, for larger $\alpha$, the spectrum cannot be fully converted during the propagation, and the Gaussian pattern is gradually replaced by the Airy pattern, as shown in Fig. 5(d3).

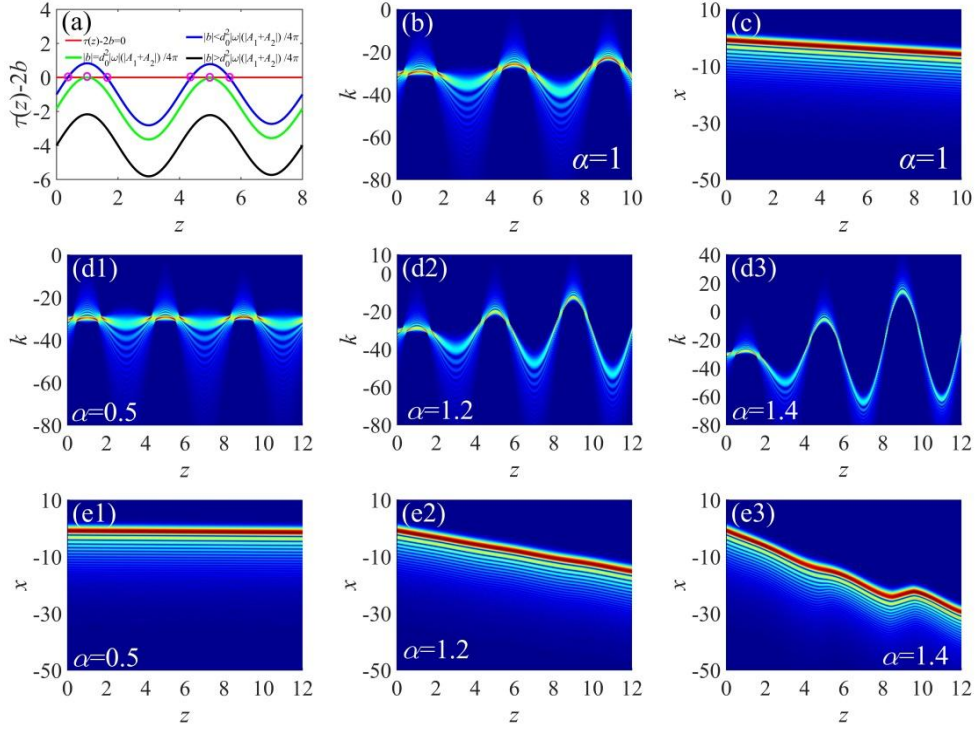

Fig. 5 (a) The constraint relation for the spectrum conversion in the case of $\tau(z) = 2b$, and evolution of the chirped Airy beam in the momentum and the coordinate spaces for chirp $b = 0.5$ and different values of $\alpha$: (b, c) $\alpha = 1$; (d1, e1) $\alpha = 0.5$; (d2, e2) $\alpha = 1.2$; (d3, e3) $\alpha = 1.4$. Other parameters are the same as in Fig. 2.

### *3.2 The HO-hyperbolic-secant potential*

In this section, we consider the potential taking the form of the hyperbolic secant in the longitudinal direction, i.e., $v(z) = d_1 \text{sech}[\sigma(z - z_d)]$, where $d_1$, $\sigma$ and $z_d$ denote the height, width and position of the potential in the longitudinal direction. Figures 6(a) and (b), respectively, present the evolution of the chirp-free Airy beam in the momentum and the coordinate spaces in the fractional system with the HO-hyperbolic-secant potential, for $\alpha = 1$. It can be observed that the singular point extends to a finite section, i.e., $[0, z_j]$, where the beam evolves, keeping the Gaussian pattern in the momentum space. For a short distance after $z_j$, because

$\tau ( z ) = d_1 \int_0^z \mathrm{sech} \left[ \sigma ( \varepsilon - z_d ) \right] d\varepsilon$ is very close to 0, the beam still travels in the form of the Gaussian in the momentum space, as demonstrated in Eq. (15a). Until $\tau(z)$ approaches close to $z_d$, the beam contracts and then expands, evolving into an apparent Airy pattern. That is, the beam undergoes tunneling through the hyperbolic-secant potential. Moreover, comparing Figs. 6(a) and (b), it is seen that the width and longitudinal central position of the hyperbolic-secant potential strongly affect the position of the compression point and the size of the singular section $[0, z_j]$ in the momentum space and also affect the splitting and propagation direction in the coordinate space. Figures 6(c) and (d) display the relationship of the compression point and $z_j$ to the width factor $\sigma$ and the center position $z_d$ of the potential, respectively, from where one can see that the increase of $\sigma$ and $z_d$ leads to the increase of the distance from the input position to the compression point and singularity position $z_j$. For the case of $\alpha \neq 1$, the variation of $\alpha$ leads to decrease or increase of the spectral shift of the converted Airy pattern in the momentum space after

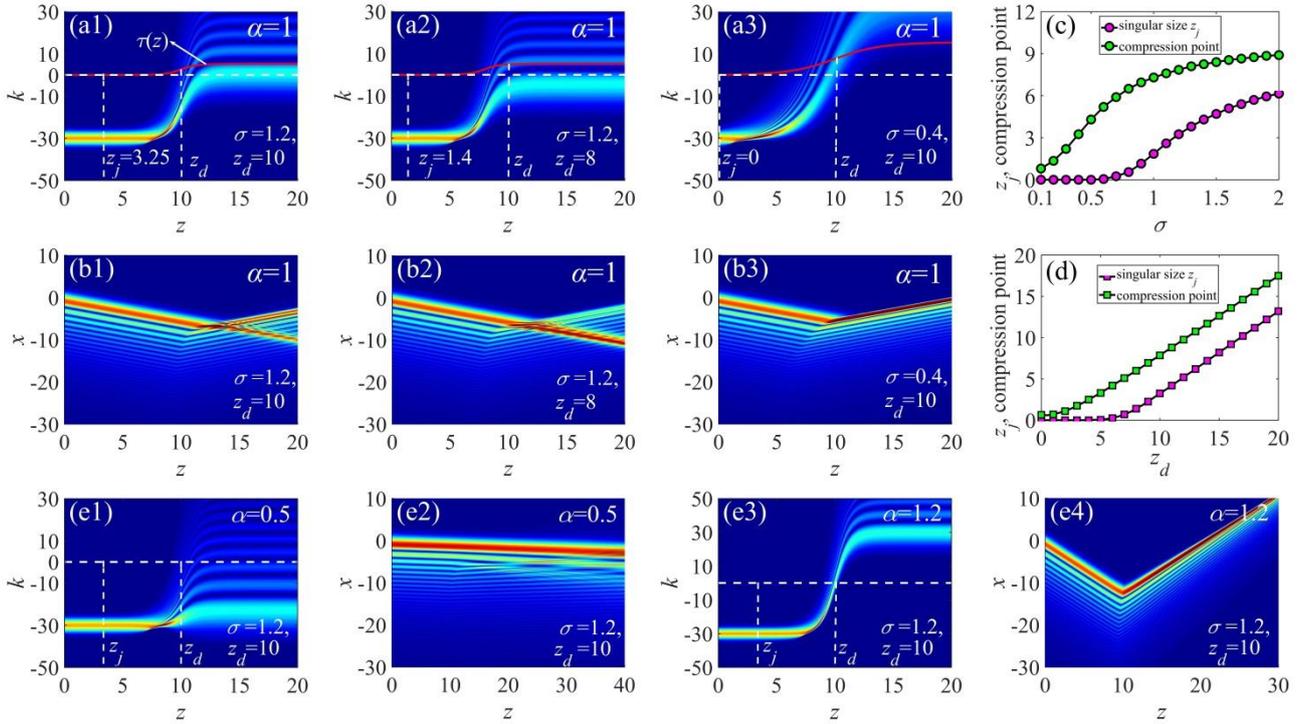

Fig. 6 (a,b) The evolution of the chirp-free Airy beam in (a) the momentum and (b) the coordinate spaces for $\alpha = 1$: (a1,b1) $\sigma = 1.2$, $z_d = 10$, (a2,b2) $\sigma = 1.2$, $z_d = 8$, (a3,b3) $\sigma = 0.4$, $z_d = 10$. (c,d) The relation of $z_j$ and the compression point with width factor $\sigma$ and center's coordinate $z_d$ for $\alpha = 1$. (e) The evolution of the chirp-free Airy beam for $\sigma = 1.2$, $z_d = 10$ and $\alpha \neq 1$: (e1,e2) $\alpha = 0.5$; (e3,e4) $\alpha = 1.2$. Other parameters are: $C = -30$, $d_1 = 2$.

passing the compression points, and dramatically affects the evolution direction of the beam in the coordinate space, which can be seen by comparing Figs. 6(a1) and (b1) with Figs. 6(e1)-(e4). These characteristics suggest a possibility to engineer the tunneling behavior of the beam by adjusting the width factor and center's position of the potential and Lévy index.

For the chirped Airy beam in the HO-hyperbolic secant potential, according to condition $\tau(z) = 2b$, one can derive the relation between chirp $b$ and parameters of the potential, which is necessary to realize the spectrum conversion between the Airy and Gaussian patterns, as

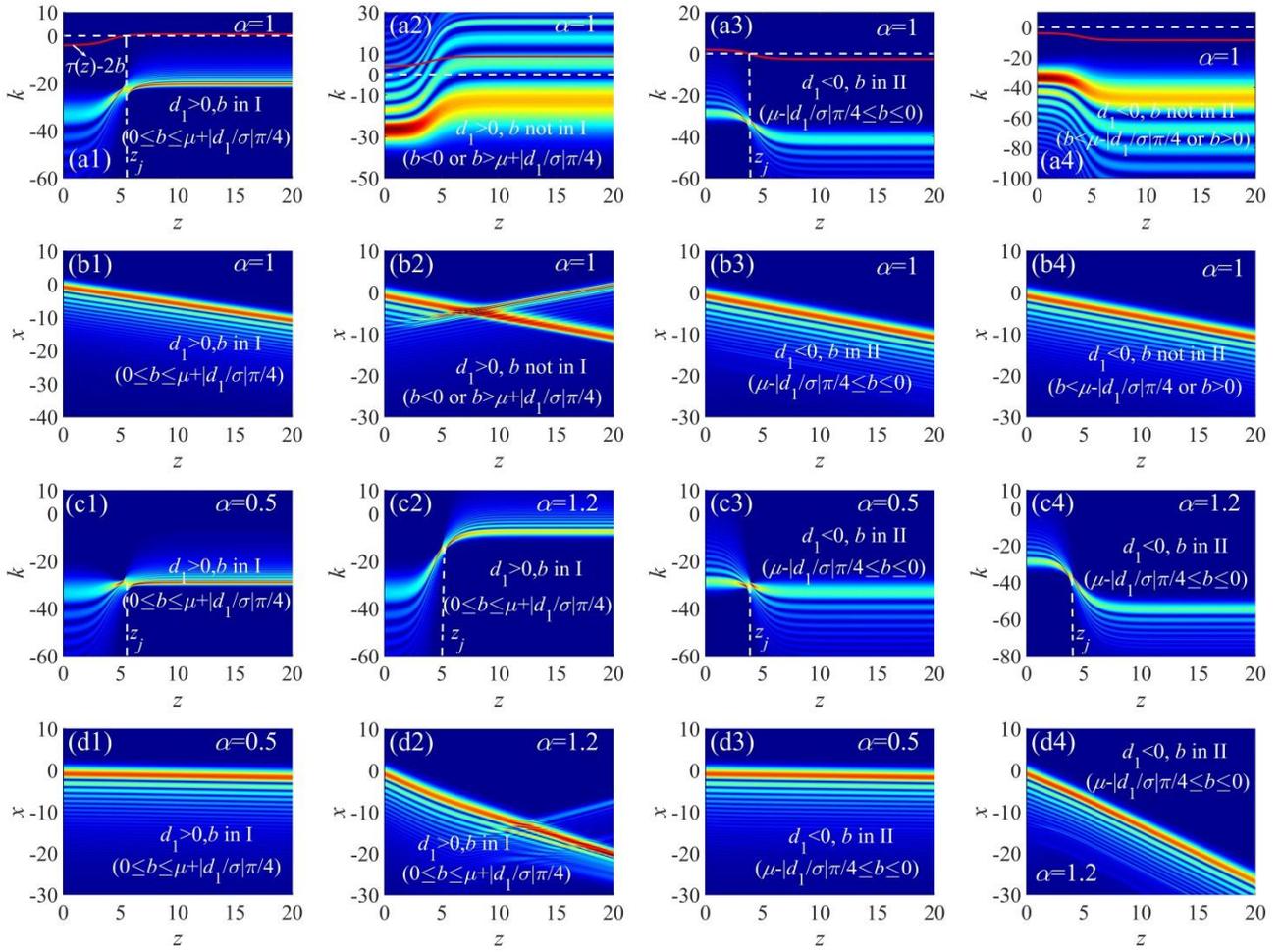

Fig. 7 The evolution of the chirped Airy beam in the momentum (the first and third rows) and the coordinate (the second and fourth rows) spaces. (a,b) For $\alpha = 1$: (a1,b1) $d_1 > 0$, $b$ in Range I; (a2,b2) $d_1 > 0$, $b$ not in Range I; (a3,b3) $d_1 < 0$, $b$ in Range II; (a2,b2) $d_1 < 0$, $b$ not in Range II. (c,d) For $\alpha \neq 1$, when $b$ lies in Range I or II: (c1,d1) $\alpha = 0.5$ and (c2,d2) $\alpha = 1.2$, when $b$ lies in Range I; (c3,d3) $\alpha = 0.5$ and (c4,d4) $\alpha = 1.2$, when $b$ lies in

Range II. The other parameter is $C=-30$ and $\mu=\dfrac{d_1 \arcsin\left[\text{th}(\sigma z_d)\right]}{2\sigma}$ in all the panels.

$$0 \leq b \leq \dfrac{d_1 \arcsin\left[\text{th}(\sigma z_d)\right]}{2\sigma} + \dfrac{\pi}{4}\left|\dfrac{d_1}{\sigma}\right| \quad \text{for} \quad d_1 > 0 \quad \text{(named Range I), and}$$

$$\dfrac{d_1 \arcsin\left[\text{th}(\sigma z_d)\right]}{2\sigma} - \dfrac{\pi}{4}\left|\dfrac{d_1}{\sigma}\right| \leq b \leq 0 \quad \text{for} \quad d_1 < 0 \quad \text{(Range II)}.$$

As shown in Figs. 7(a1) and (a3), the chirped Airy beam undergoes tunneling and spectrum conversion at singular position $z_j$ when chirp $b$ lies within the Ranges I or II; while the chirped Airy beam exhibits the Airy-pattern preservation when chirp $b$ does not satisfy the above-mentioned relation, as depicted in Figs. 7(a2) and (a4). On the other hand, the chirped Airy beam keeps the Airy pattern propagating in the coordinate space, regardless of $b$ and the potential parameters, as displayed in Figs. 7(b). However, chirp $b$ and the potential parameters markedly affect the split and evolution direction of the beam in the coordinate space, as they affect the motion of the Airy lobes across the axis of $k=0$ in the momentum space. Moreover, the evolution direction and splitting of the chirped Airy beam also depend on the Lévy index $\alpha$ in the case of the HO-hyperbolic-secant potential, as shown in Fig. 7 (d). Accordingly, in the momentum space, $\alpha$ produces a significant effect on the spectrum evolution. Figure 7(c) presents the spectral characteristics of the Airy beam, with chirp $b$ lying in Range I or Range II for different $\alpha$. It is seen that the spectrum conversion still happens at point $z=z_j$ for $\alpha<1$ and $\alpha>1$, and the increase of $\alpha$ causes the shift of the spectrum along the positive $k$ direction when $b$ lies within Range I [see Figs. 7(c1) and (c2)], or along the negative $k$ direction when $b$ lies in Range II [see Figs. 7(c3) and (c4)].

### 3.3 The HO-delta-functional potential

As a natural extension of the above analysis, we consider the longitudinal modulation in the form of the Dirac's delta-function, $v(z)=d_2\delta(z-z_\delta)$, where $d_2$ and $z_\delta$ are the height and position of the potential. In this case, it is easy to obtain $\tau(z)=d_2 H(z-z_\delta)$ and $\varsigma(z)=d_2 z_\delta H(z-z_\delta)$, in which $H(\bullet)$ denotes the Heaviside step function. According to Eqs. (15a) and (15b), one can find that the abrupt spectrum conversion between the Gaussian and Airy patterns occurs at $z_j=z_\delta$ in the momentum space [see Figs. 8(a1) and (a2)], and the Airy pattern with unchanged velocity is perservered in the coordinate space [see Figs. 8(b1) and (b2)]. Inset of Fig. 8(a1) demonstrates that the Airy pattern at $z = z_\delta$ undergoes a two-step spectrum shift due to the double jump of $\tau(z)$ and $\varsigma(z)$ at $z = z_\delta$ [38], which is also explicitly presented in Figs. 8(c) and (d). Moreover, comparing the evolution of the beams shown in Figs. 8(a1) and (a2), one finds that the Airy patterns are reversed along the $k$ direction for $d_2 > 0$ and $d_2 < 0$, which is attributed to the opposite sign of $\tau(z)$ at $z > z_\delta$. Figures 8(e) and (f) clearly display the effect of $\alpha$ on the evolution of the beam in the momentum and the coordinate spaces. It is observed that larger Lévy index $\alpha$ causes a more apparent spectrum shift of the Airy pattern at $z > z_\delta$ in the momentum space and faster velocity of the beam motion in the coordinate space.

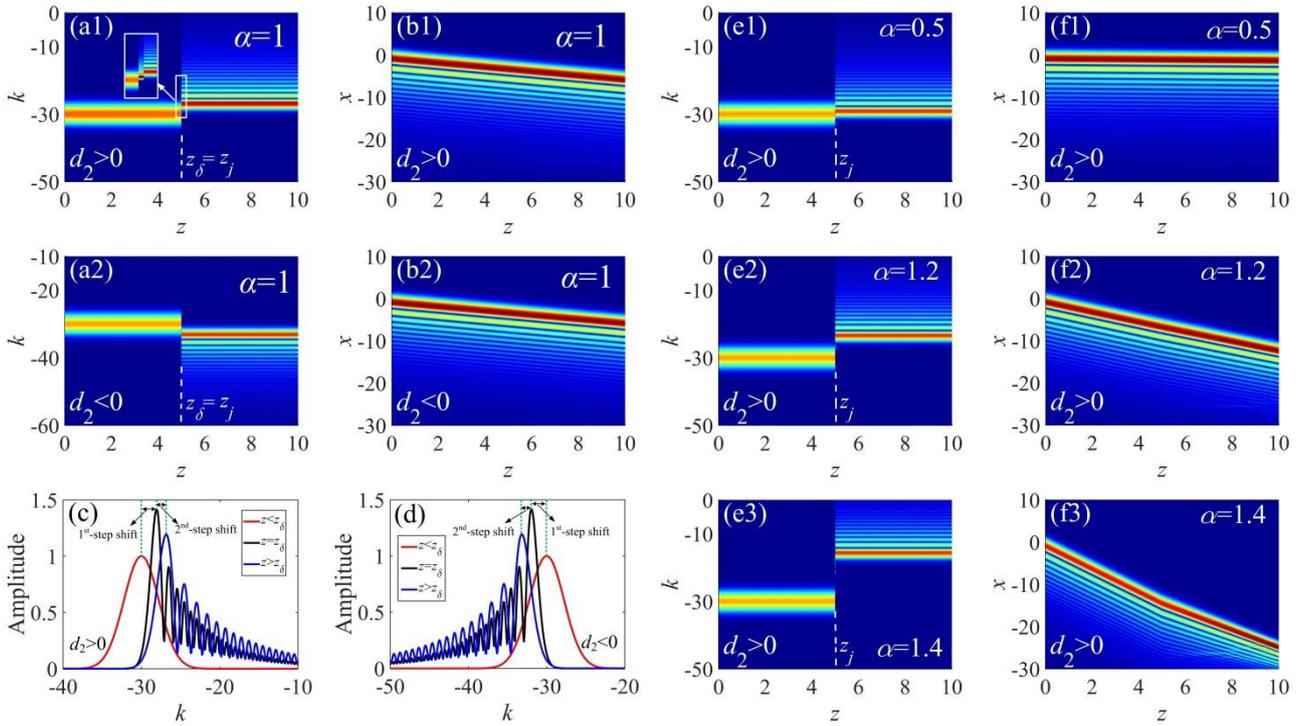

Fig. 8 (a,b) The evolution of the chirp-free Airy beam in (a) the momentum and (b) the coordinate spaces for $\alpha = 1$: (a1,b1) $d_2 > 0$; (a2,b2) $d_2 < 0$ and (c,d) the spectrum shifts for $d_2 > 0$ and $d_2 < 0$, corresponding to (a1,a2); (e,f) the evolution of the chirp-free Airy beam for $\alpha \neq 1$ when $d_2 > 0$: (e1,f1) $\alpha = 0.5$, (e2,f2) $\alpha = 1.2$, and (e3,f3) $\alpha = 1.4$. Other parameters are $C = -30$, $z_\delta = 5$.

Finally, we investigate the evolution of the chirped Airy under the action of the delta-functional barrier potential (i.e., $d_2 > 0$), which are shown in Fig. 9. In agreement with Eqs. (18a) and (18b), the beam keeps the Airy pattern in both the momentum and coordinate spaces when $b \neq d_2/2$, as shown in Figs. 9 (a1,b1) and (a2,b2). Moreover, an abrupt spectrum shift appears at $z = z_\delta$. Note that the spectrum's energy at $z < z_\delta$ is different from that at $z > z_\delta$, and the energy is also different for $b < 0$ and $b > 0$ [see Figs. 9(a1) and (a2)], which can be explained from by amplitude $\left|\sqrt{2\pi/\left[d_2 H(z - z_\delta) - 2b\right]}\right|$ in Eq. (18a). For the delta-functional barrier potential ($d_2 > 0$), with $b < 0$ in Fig. 9(a1), the amplitude of the Airy pattern after the spectrum shift is smaller than before the shift. On the other hand, for $b > 0$ in Fig. 9(a2), the change of the spectrum's energy is opposite. Further, when $b = d_2/2$, the spectrum conversion between the Airy and Gaussian patterns occurs at $z = z_\delta$ in the momentum space [see Fig. 9(a3)] due to relation $\tau(z) - 2b = -2b H(z_\delta - z)$, as demonstrated by Eqs. 18(a) and 19(a). As expected, in the coordinate space, the chirped Airy beam exhibits the Airy pattern preservation, as depicted in Fig. 9(b3). As for the effect of $\alpha$ on the

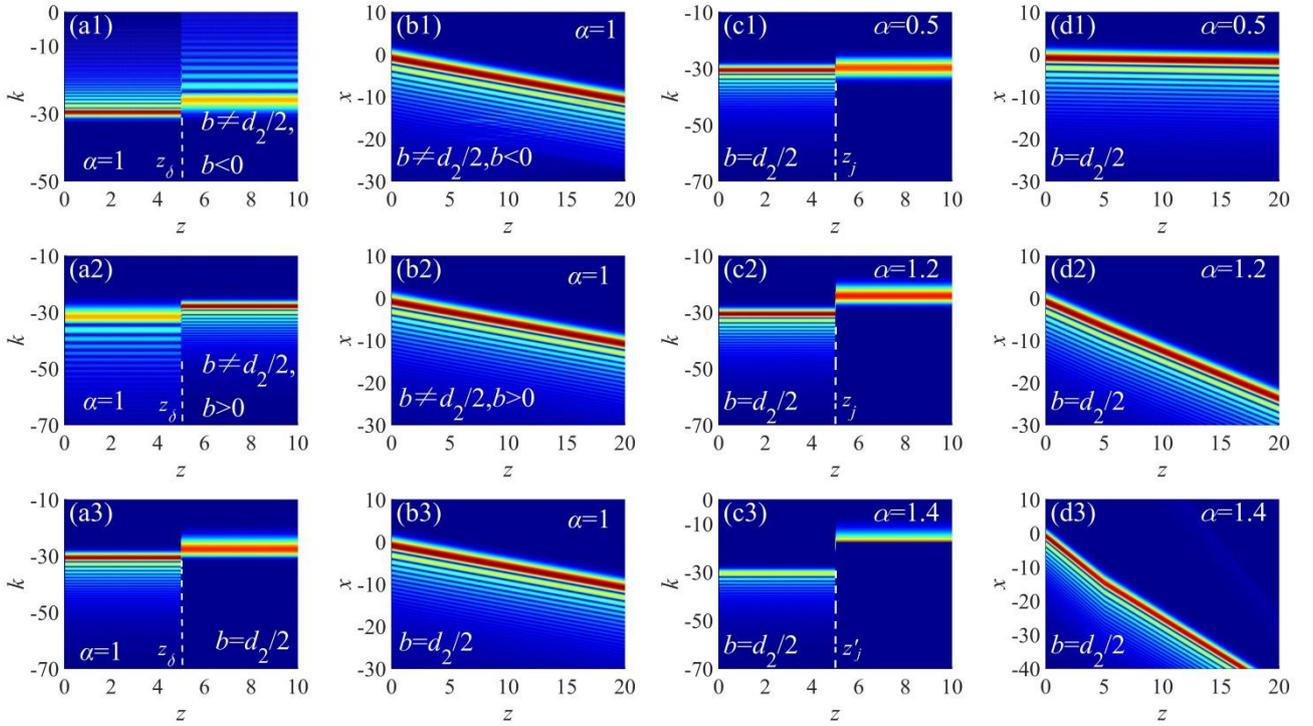

Fig. 9 The evolution of the chirped Airy beam in the momentum (the first and third columns) and the coordinate (the second and fourth columns) spaces. (a,b) For $\alpha = 1$: (a1,b1) $b \neq d_2/2$, $b < 0$, (a2,b2) $b \neq d_2/2$, $b > 0$, (a3,b3)

$b = d_2/2$; (c,d) for $\alpha \neq 1$ and $b = d_2/2$: (c1,d1) $\alpha = 0.5$, (c2,d2) $\alpha = 1.2$, (c3,d3) $\alpha = 1.4$. Other parameters are the same as in Fig. 8 (a1).

chirped Airy beam in the case of the delta-functional barrier potential, it is concluded from the results shown in Figs. 9(c) and (d) that the spectrum shift and evolution velocity feature a consistent difference from those of the chirp-free Airy beam, as displayed in Fig. 8. However, for larger $\alpha$, the spectrum conversion between the Airy and Gaussian pattern does not take place, as seen in Fig. 9(c3). Accordingly, the propagation distance marked by the white dot line is labeled as $z'_j$ in Fig. 9(c3) to exhibit the effect of $\alpha$ on the spectrum conversion.

## 4. Conclusion

In this work we have studied the evolution characteristics of Airy beams in the framework of the linear one-dimensional FSE (fractional Schrödinger equation) with the modulated HO (harmonic-oscillator) potential. First, we derived the general analytical solutions describing the evolution of beams in the fractional system, and presented particular solutions for the chirp-free and chirped Airy beams, in both the momentum and coordinate spaces. Then, based on these solutions, we explored the spectrum conversion and pattern preservation of the Airy beams with/without chirp in HO-moiré-lattice, HO-hyperbolic-secant and HO-delta-functional potentials. Under the action of the HO-moiré-lattice potential, the chirp-free Airy beam undergoes multiple spectrum conversions between the Airy and Gaussian patterns in the momentum space, and preserves the Airy pattern in the coordinate space. The chirp-free Airy beam seems to experience the action of an effective restoring force in the course of the propagation. Moreover, parameters of the modulated moiré lattice affect the singular positions of the spectrum conversion and oscillation amplitudes of the Airy lobes. In that case, the chirped Airy beam preserves the Airy patterns in both the momentum and coordinate spaces in the case of $\tau(z) \neq 2b$ (recall that $b$ is the chirp parameter, and $\tau(z)$ is determined by the integration of the modulated moiré lattice along the propagation distance). On the other hand, in the case of $\tau(z) = 2b$ the spectrum conversion occurs only when the chirp satisfies a constraint condition. For the HO-hyperbolic-secant potential, the chirp-free Airy beam experiences a spectrum conversion and tunneling. Specifically, within the singular section the beam travels in the form of the Gaussian; beyond the singular section, it gradually contracts and then expands, evolving into an apparent Airy pattern. The spectrum conversion and pattern preservation of the chirped Airy beam are determined by the chirp and parameters of the hyperbolic-secant potential barrier. Under the action of the HO-delta-functional potential, the chip-free Airy beam undergoes the abrupt spectrum conversion and two-step spectrum shift of the Airy pattern. For the chirped Airy beam, it is found that the Airy pattern can be kept in the momentum and the coordinate spaces alike, when $b \neq d_2/2$ (recall that $d_2$ is the height of the delta-functional potential), while the spectrum conservation occurs only when the chirp satisfies relation $b = d_2/2$. The Lévy index is found to strongly affect the spectrum conversion and pattern preservation. These results may be used for the design of "bespoken" optical beams and optical modulators. As an extension of the present analysis, it may be relevant to consider similar dynamical scenarios in the nonlinear FSE. As well as in its two-dimensional version.

## Acknowledgments

This work was supported by the National Natural Science Foundation of China (Grant numbers 61775126, 62305199); the Natural Science Foundation of Shanxi Province (Grant number 202203021221016); the Israel Science Foundation (Grant number 1695/22).

## Data availability

Data may be made available upon a request.

## Declaration of competing interest

The authors declare that they have no known competing financial interests or personal relationships that could have appeared to influence the work reported in this paper.

## CRediT authorship contribution statement

**Xiaoqin Bai:** Methodology, Formal analysis, Writing-original draft, Data curation. **Juan Bai:** Software, Writing-review & editing. **Boris A. Malomed**: Conceptualization, Formal analysis, Drafting and editing the paper. **Rongcao Yang:** Conceptualization, Methodology, Resources, Supervision, Funding acquisition.

## References

[1] Zhang SH, Wang C, Zhang HL, Ma P, Li XK. Dynamic analysis and bursting oscillation control

of fractional- order permanent magnet synchronous motor system. Chaos Soliton Fractal 2022; 156:111809.

[2] Laskin N. Fractional quantum mechanics and Lévy path integrals. Phys Lett A 2000; 268(4–6): 298–305.

[3] Laskin N. Fractional quantum mechanics. Phys Rev E 2000; 62(3): 3135–3145.

[4] Laskin N. Fractional Schrödinger equation. Phys Rev E 2002; 66(5): 056108.

[5] Longhi S. Fractional Schrödinger equation in optics. Opt Lett 2015; 40(6):1117–1120.

[6] Liu SL, Zhang YW, Malomed BA, Karimi E. Experimental the realisations of the fractional Schrödinger equation in the temporal domain. Nat Commun 2023; 14(1): 222.

[7] Zhang YQ, Zhong H, Belić, MR, Ahmed N, Zhang YP, Xiao M. Diffraction-free beams in fractional Schrödinger equation. Sci Rep 2016; 6: 23645.

[8] Zhang YQ, Liu X, Belić MR, Zhong WP, Zhang YP, Xiao M. Propagation dynamics of a light beam in a fractional Schrödinger equation. Phys Rev Lett 2015; 115(18):180403.

[9] Zhang YQ, Wang R, Zhong H, Zhang JW, Belić MR, Zhang YP. Optical Bloch oscillation and Zener tunneling in the fractional Schrödinger equation. Sci Rep 2017; 7: 17827.

[10] Zhang YQ, Wang R, Zhong H, Zhang JW, Belić MR, Zhang YP. Resonant mode conversions and Rabi oscillations in a fractional Schrödinger equation. Opt Express 2017; 25(26): 32401–32410.

[11] Huang CM, Shang C, Li J, Dong LW, Ye FW. Localization and Anderson delocalization of light in fractional dimensions with quasi-periodic lattice. Opt Express 2019; 27(5): 6259–6267.

[12] Li PF, Malomed BA, Mihalache D. Vortex solitons in fractional nonlinear Schrödinger equation with the cubic-quintic nonlinearity. Chaos Soliton Fractal 2020; 137: 109783.

[13] Malomed BA. Optical Solitons and Vortices in Fractional Media: A Mini-Review of Recent Results. Photonics. 2021; 8(9): 353.

[14] Li PF, Sakaguchi H, Zeng LW, Zhu X, Mihalache D, Malomed BA. Second-harmonic generation in the system with fractional diffraction. Chaos Soliton Fractal 2023; 173: 113701.

[15] Malomed BA. Basic fractional nonlinear-wave models and solitons. Chaos 2024; 34, 022102.

[16] Zhang LF, Li CX, Zhong HZ, Xu CW, Lei DJ, Li Y, Fan DY. Propagation dynamics of super-Gaussian beams in fractional Schrödinger equation: from linear to nonlinear regimes. Opt Express 2016; 24(13): 14406-14418.

[17] Dong LW, Liu DS, Qi W, Wang LX, Zhou H, Peng P, Huang CM. Necklace beams carrying

fractional angular momentum in fractional systems with a saturable nonlinearity. Commun Nonlinear Sci 2021; 99:105840.

[18] He SL, Peng X, He YJ, Deng DM. Autofocus properties of astigmatic chirped symmetric Pearcey Gaussian vortex beams in the fractional Schrödinger equation with parabolic potential. Opt Express 2023; 31(11): 17930-17942.

[19] Tehrani DHT, Solaimani M. Effects of the medium fractionality and oscillating potential profiles on the Superarrivals of the Gaussian wave packets. Chaos Soliton & Fractals 2023; 168:113138.

[20] Berry MV, Balazs NL. Nonspreading wave packets. Am J Phys 1979; 47(3): 264.

[21] Christodoulides DN, Coskun TH. Diffraction-free planar beams in unbiased photorefractive media. Opt Lett 1996; 21(18): 1460.

[22] Broky J, Siviloglou GA, Dogariu A, and Christodoulides DN. Self-healing properties of optical Airy beams. Opt Express 2008; 16(17): 12880–12891.

[23] Chen WJ, Wang T, Wang J, Mu YN. Dynamics of interacting Airy beams in the fractional Schrödinger equation with a linear potential. Opt Commun 2021; 496: 127136.

[24] Iomin A. Fractional Schrodinger equation in gravitational optics. Mod Phys Lett A 2021; 36(14): 2140003.

[25] Bai XQ, Yang RC, Jia HP, Bai J. Dynamics and manipulation of Airy beam in fractional system with diffraction modulation and PT-symmetric potential. Nonlinear Dyn 2023; 111(5): 4577-4591.

[26] Liang ZX, Zhang ZD, Liu WM. Dynamics of a bright soliton in Bose-Einstein condensates with time-dependent atomic scattering length in an expulsive parabolic potential. Phys Rev Lett 2005; 94(5): 050402.

[27] Zhan KY, Zhang WQ, Jiao RY, Liu B. Period-reversal accelerating self-imaging and multi-beams interference based on accelerating beams in parabolic optical potentials. Opt Express 2020; 28(14): 20007-20015.

[28] Wang Q, Mihalache D, Belić MR, Zeng LW, Lin J. Spiraling Laguerre–Gaussian solitons and arrays in parabolic potential wells. Opt Letters 2023; 48(16): 4233-4236.

[29] Snyder AW, Mitchell DJ. Accessible Solitons. Science 1997; 276 (5318): 1538-1541.

[30] Zhang YQ, Liu X, Belić MR, Zhong WP, Wen F, Zhang YP. Anharmonic propagation of two-dimensional beams carrying orbital angular momentum in a harmonic potential. Opt Letters 2015; 40 (16): 3786-3789.

[31] Efremidis NK. Airy trajectory engineering in dynamic linear index potentials. Opt Lett 2011; 36(15): 3006–3008.

[32] Bernardini C, Gori F, Santarsiero M. Converting states of a particle under uniform or elastic forces into free particle states. Eur J Phys 1995; 16: 58-62.

[33] Zhang LF, Liu K, Zhong HZ, Zhang JG, Li Y, Fan DY. Effect of initial frequency chirp on Airy pulse propagation in an optical fiber. Opt Express 2015; 23(3): 2566-2576.

[34] Wang P, Zheng YL, Chen XF, Huang CM, Kartashov YV, Torner L, Konotop VV, Ye FW. Localization and delocalization of light in photonic moiré lattices. Nature 2020; 577(7788): 42-46.

[35] Wang P, Fu QD, Peng RH, Kartashov YV, Torner L, Konotop VV, Ye FW. Two-dimensional Thouless pumping of light in photonic moiré lattices. Nat Commun 2022; 13(1): 6738.

[36] Xue RD, Wang W, Wang LQ, Chen HL, Guo RP, Chen J. Localization and oscillation of optical beams in Moiré lattices. Opt Express 2017; 25(5): 5788-5796.

[37] Talukdar TH, Hardison AL, Ryckman JD. Moiré Effects in Silicon Photonic Nanowires. ACS Photonics. 2022; 9(4): 1286-1294.

[38] Karjanto N, Hanif W, Malomed BA, Susanto H. Interactions of bright and dark solitons with localized *PT*-symmetric potentials. Chaos 2015; 25: 023112.